\documentclass[11pt]{article}

\usepackage[margin=1in]{geometry}
\usepackage[T1]{fontenc}
\usepackage[utf8]{inputenc}
\usepackage{lmodern}
\usepackage{microtype}
\usepackage{setspace}

\usepackage{authblk}

\usepackage{amsmath,amssymb}
\usepackage{siunitx}
\usepackage{graphicx}
\graphicspath{{./}}
\usepackage{booktabs}
\usepackage{array}
\usepackage{tabularx}
\usepackage{longtable}
\usepackage{adjustbox}
\usepackage{multirow}
\usepackage{float}
\usepackage{placeins}
\usepackage{caption}
\usepackage{subcaption}
\usepackage{tikz}
\usetikzlibrary{arrows.meta,positioning,shapes.geometric,fit,calc}

\usepackage{enumitem}
\setlist[itemize]{leftmargin=1.5em}
\setlist[enumerate]{leftmargin=1.5em}

\usepackage{xcolor}
\usepackage{xurl}
\usepackage[numbers,sort&compress]{natbib}
\usepackage{hyperref}
\usepackage[nameinlink,noabbrev]{cleveref}

\hypersetup{
    colorlinks=true,
    linkcolor=blue!50!black,
    citecolor=blue!50!black,
    urlcolor=blue!50!black,
    pdftitle={Nonlinear MESA-RSP Modeling of Delta Cephei: Period Matching, Amplitude Control, and Model-Acceptance Diagnostics},
    pdfauthor={Zuhoor Elahi, Christopher Sirola, Wafa Gull}
}

\graphicspath{{./}{}}

\newcommand{\safeincludegraphics}[3][width=0.92\linewidth]{%
  \IfFileExists{#2}{%
    \includegraphics[#1]{#2}%
  }{%
    \fbox{%
      \begin{minipage}[c][0.26\textheight][c]{0.88\linewidth}
      \centering
      \vspace{0.5em}
      \textbf{Missing figure file}\par
      \vspace{0.5em}
      #3\par
      \vspace{0.5em}
      \texttt{\detokenize{#2}}\par
      \vspace{0.5em}
      Place the figure file at this path and recompile.
      \end{minipage}}%
  }%
}

\newcommand{\safetableinput}[1]{%
  \IfFileExists{#1}{%
    \input{#1}%
  }{%
    \fbox{%
      \begin{minipage}{0.92\linewidth}
      \centering
      \vspace{0.8em}
      \textbf{Missing table file}\par
      \vspace{0.5em}
      \texttt{\detokenize{#1}}\par
      \vspace{0.8em}
      \end{minipage}}%
  }%
}

\newcommand{\dcep}{\ensuremath{\delta} Cephei}
\newcommand{\Msun}{\ensuremath{M_{\odot}}}

\newcommand{\Lsun}{\ensuremath{L_{\odot}}}
\newcommand{\Teff}{\ensuremath{T_{\rm eff}}}

\newcommand{\MESA}{\textsc{MESA}}
\newcommand{\RSP}{\textsc{RSP}}
\newcommand{\MESARSP}{\MESA-\RSP}

\newcommand{\code}[1]{\texttt{\detokenize{#1}}}

\newcommand{\RSPalfam}{\ifmmode\mathrm{RSP\_alfam}\else\texttt{RSP\_alfam}\fi}
\newcommand{\RSPgammar}{\ifmmode\mathrm{RSP\_gammar}\else\texttt{RSP\_gammar}\fi}
\newcommand{\RSPalfat}{\ifmmode\mathrm{RSP\_alfat}\else\texttt{RSP\_alfat}\fi}

\newcommand{\workflowfigure}{%
\begin{figure}[p]
\centering
\resizebox{0.92\textwidth}{!}{%
\begin{tikzpicture}[
font=\small,
node distance=6mm and 8mm,
box/.style={draw, rounded corners, align=center, minimum width=4.3cm, minimum height=0.82cm, fill=blue!6},
smallbox/.style={draw, rounded corners, align=center, minimum width=4.3cm, minimum height=0.82cm, fill=green!6},
decision/.style={draw, diamond, aspect=2.25, align=center, inner xsep=1.5ex, fill=orange!10},
arrow/.style={-{Latex[length=2.2mm]}, thick}
]
\node[box] (setup) {Delta Cephei active/header inlists};
\node[decision, below=of setup] (verify) {Verified active\\inlist/header?};
\node[box, right=of verify] (fix) {Correct run linkage\\and re-check};
\node[box, below=of verify] (grid) {Initial stellar setup and trial runs};
\node[box, below=of grid] (period) {Period calibration\\(200 periods)};
\node[decision, below=of period] (pmatch) {$P_{\rm model}$ near\\$P_{\rm obs}$?};
\node[box, right=of pmatch] (retune) {Retune stellar parameters\\or setup choices};
\node[box, below=of pmatch] (extend) {Extend nonlinear run\\and inspect behavior};
\node[box, below=of extend] (alfam) {Amplitude-control sequence in $\RSPalfam$};
\node[box, below=of alfam] (diag) {Cycle diagnostics:\\$P_n$, $\Delta R_n$, $\max(v_{\rm surf}/c_s)$, warning flags};
\node[decision, below=of diag] (accept) {Clean late-cycle\\behavior and no\\serious warning?};
\node[smallbox, below left=8mm and 4mm of accept] (clean) {Clean / reference /\\amplitude-enhanced candidate};
\node[smallbox, below right=8mm and 4mm of accept] (diagnostic) {Diagnostic or rejected model};
\draw[arrow] (setup) -- (verify);
\draw[arrow] (verify) -- node[left, font=\scriptsize]{yes} (grid);
\draw[arrow] (verify) -- node[above, font=\scriptsize]{no} (fix);
\draw[arrow] (fix.north) |- (setup.east);
\draw[arrow] (grid) -- (period);
\draw[arrow] (period) -- (pmatch);
\draw[arrow] (pmatch) -- node[left, font=\scriptsize]{yes} (extend);
\draw[arrow] (pmatch) -- node[above, font=\scriptsize]{no} (retune);
\draw[arrow] (retune.north) |- (grid.east);
\draw[arrow] (extend) -- (alfam);
\draw[arrow] (alfam) -- (diag);
\draw[arrow] (diag) -- (accept);
\draw[arrow] (accept) -- node[above left, font=\scriptsize]{yes} (clean);
\draw[arrow] (accept) -- node[above right, font=\scriptsize]{no} (diagnostic);
\end{tikzpicture}%
}
\caption{Reproducible workflow used in this paper. The active Delta Cephei inlist/header must first be verified. Period calibration is then followed by nonlinear extension, amplitude-control tests, cycle diagnostics, and final clean-versus-diagnostic model classification.}
\label{fig:rsp_workflow}
\end{figure}
}

\newcommand{\classificationflowchart}{%
\begin{figure}[htbp]
\centering
\resizebox{0.90\textwidth}{!}{%
\begin{tikzpicture}[
font=\small,
node distance=8mm and 7mm,
box/.style={draw, rounded corners, align=center, minimum width=3.0cm, minimum height=0.9cm, fill=blue!6},
outcome/.style={draw, rounded corners, align=center, minimum width=3.2cm, minimum height=0.95cm},
decision/.style={draw, diamond, aspect=2.0, align=center, inner xsep=1.4ex, fill=orange!10},
arrow/.style={-{Latex[length=2.2mm]}, thick}
]
\node[box] (run) {Completed model run};
\node[decision, below=of run] (setupok) {Correct setup and\\usable output?};
\node[outcome, fill=red!8, right=of setupok] (reject1) {Rejected\\(wrong setup or unusable output)};
\node[decision, below=of setupok] (periodok) {Period acceptably\\close to target?};
\node[outcome, fill=red!8, right=of periodok] (reject2) {Rejected\\(poor period behavior)};
\node[decision, below=of periodok] (behaveok) {Interpretable late-cycle\\amplitude behavior?};
\node[outcome, fill=yellow!12, right=of behaveok] (diag1) {Diagnostic\\(nonstationary or incomplete saturation)};
\node[decision, below=of behaveok] (warnok) {Serious energy or\\numerical warning?};
\node[outcome, fill=yellow!12, right=of warnok] (diag2) {Diagnostic only\\(informative but not adopted)};
\node[decision, below=of warnok] (ampenh) {Baseline-like or\\enhanced clean model?};
\node[outcome, fill=green!12, below left=8mm and 6mm of ampenh] (ref) {Reference / clean\\baseline model};
\node[outcome, fill=green!18, below right=8mm and 6mm of ampenh] (cand) {Clean amplitude-enhanced\\candidate};
\draw[arrow] (run) -- (setupok);
\draw[arrow] (setupok) -- node[left, font=\scriptsize]{yes} (periodok);
\draw[arrow] (setupok) -- node[above, font=\scriptsize]{no} (reject1);
\draw[arrow] (periodok) -- node[left, font=\scriptsize]{yes} (behaveok);
\draw[arrow] (periodok) -- node[above, font=\scriptsize]{no} (reject2);
\draw[arrow] (behaveok) -- node[left, font=\scriptsize]{yes} (warnok);
\draw[arrow] (behaveok) -- node[above, font=\scriptsize]{no} (diag1);
\draw[arrow] (warnok) -- node[left, font=\scriptsize]{no} (ampenh);
\draw[arrow] (warnok) -- node[above, font=\scriptsize]{yes} (diag2);
\draw[arrow] (ampenh) -- node[above left, font=\scriptsize]{baseline} (ref);
\draw[arrow] (ampenh) -- node[above right, font=\scriptsize]{enhanced} (cand);
\end{tikzpicture}%
}
\caption{Model-classification logic used in this paper. A model is accepted as clean only if the setup is correct, the period remains close to the target, the late-cycle behavior is interpretable, and no serious numerical warning is present. Scientifically informative warning-producing runs are retained as diagnostic rather than over-interpreted as final solutions.}
\label{fig:model_classification_flowchart}
\end{figure}
}

\newcommand{\periodcalibrationfigure}{%
\begin{figure}[htbp]
\centering
\begin{tikzpicture}[
scale=1.0,
box/.style={draw, rounded corners, align=center, minimum width=3.4cm, minimum height=1.0cm, fill=blue!6},
smallbox/.style={draw, rounded corners, align=center, minimum width=3.2cm, minimum height=0.9cm, fill=green!8},
arrow/.style={-{Latex[length=2.5mm]}, thick}
]
\node[box] (obs) at (0,0) {Observed period\\$P_{\rm obs}=5.366531~{\rm d}$};
\node[box] (mod) at (6.0,0) {Tuned 200-period model\\$P_{\rm model}=5.366622~{\rm d}$};
\draw[arrow] (obs) -- node[above, align=center, font=\small] {period\\comparison} (mod);
\node[smallbox] (diff) at (3.0,-1.7) {$\Delta P=9.1\times10^{-5}~{\rm d}$\\$\simeq 7.86~{\rm s}$\\fractional error $=1.70\times10^{-5}$};
\draw[arrow] (obs.south) -- (diff.west);
\draw[arrow] (mod.south) -- (diff.east);
\end{tikzpicture}
\caption{Period calibration summary for the tuned 200-period Delta Cephei \MESARSP{} model. The agreement is excellent in period, but later nonlinear diagnostics show that period matching alone is not a sufficient acceptance criterion.}
\label{fig:period_calibration_200}
\end{figure}
}

\title{Nonlinear MESA-RSP Modeling of Delta Cephei: Period Matching, Amplitude Control, and Model-Acceptance Diagnostics}

\author[1,2,*]{Zuhoor Elahi}
\author[1]{Christopher Sirola}
\author[1]{Wafa Gull}

\affil[1]{Department of Physics and Astronomy, University of Southern Mississippi, Hattiesburg, MS, USA}
\affil[2]{Department of Physics, University of Karachi, Karachi, Pakistan}
\affil[*]{Corresponding author: zuhoor.elahi@usm.edu}

\date{}

\begin{document}
\maketitle

\begin{abstract}
Delta Cephei is the prototype classical Cepheid and a useful target for testing whether a one-dimensional nonlinear radial pulsation workflow can satisfy more than a period constraint. We present a target-specific, reproducible workflow using the Modules for Experiments in Stellar Astrophysics (\MESA{}) Radial Stellar Pulsation (\RSP{}) capability, hereafter \MESARSP{}, for constructing, tuning, extending, and classifying nonlinear radial pulsation calculations for Delta Cephei. The manuscript is deliberately framed as a controlled case study rather than as a first application of \MESARSP{} to Cepheids. A tuned 200-period calculation with stellar mass \(M=5.0~\Msun\), effective temperature \(\Teff=6050~{\rm K}\), luminosity \(L=2422.5~\Lsun\), hydrogen mass fraction \(X=0.73\), and metallicity \(Z=0.007\) gives \(P_{\rm model}=5.366622~{\rm d}\) for the adopted target \(P_{\rm obs}=5.366531~{\rm d}\), a difference of \(9.1\times10^{-5}~{\rm d}\), or approximately 7.9 s. This period agreement shows that the mean-density structure can be tuned successfully, but the extended calculation demonstrates that period matching alone is not an adequate nonlinear model-acceptance criterion. When the period-matched model was continued with the default/weak-damping setup, the amplitude and cycle behavior became nonstationary. We therefore explored the \MESARSP{} eddy-viscous damping control parameter \RSPalfam{} as a first amplitude-control experiment. The \(\RSPalfam=0.60\) case is retained as a period-stable reference, while the \(\RSPalfam=0.425\), 500-period calculation is classified as the clean amplitude-enhanced candidate from this initial sequence. A 700-period continuation is retained only as a diagnostic case because it produced an energy-error warning. The main contribution is a reproducible Delta-Cephei workflow and a transparent classification scheme showing that acceptable nonlinear Cepheid models require period agreement, interpretable late-cycle amplitude behavior, controlled surface-velocity diagnostics, and the absence of serious numerical warnings. Synthetic observed-band photometry and detailed Fourier morphology tests are treated in companion analyses.
\end{abstract}


\section{Introduction}
\label{sec:introduction}

Classical Cepheids are radially pulsating evolved stars whose periods are tightly connected to their luminosities.  They are central to stellar pulsation theory and to the extragalactic distance scale, while also serving as probes of intermediate-mass stellar evolution \citep[e.g.,][]{Cox1980,Bono1999,Bono2024}.  Delta Cephei is the prototype member of this class and is therefore a natural target for testing whether a stellar pulsation calculation can reproduce the basic global properties and nonlinear behavior of a real Cepheid.

A useful model of Delta Cephei must satisfy more than a single scalar constraint.  The observed pulsation period is the most direct and best-constrained first target, but nonlinear Cepheid modeling also involves amplitude growth, mode saturation, cycle-to-cycle stability, shock and surface-velocity behavior, and light-curve morphology.  In particular, the same approximate period can be obtained from models that behave very differently once their nonlinear amplitudes are followed for hundreds of cycles.  This distinction is important because the radial pulsation period is controlled primarily by the star's mean-density structure, whereas nonlinear amplitude and morphology depend sensitively on the adopted convection and damping treatment.

The present paper focuses on the first modeling stage of a larger Delta Cephei dissertation project: the construction of a reproducible nonlinear radial pulsation workflow using \MESARSP{}.  The \MESARSP{} framework is described in the \MESA{} instrument papers and is designed for nonlinear radial pulsation calculations of classical pulsating variables \citep{Paxton2019,Jermyn2023}.  Its physical basis connects to a long tradition of nonlinear stellar pulsation and time-dependent convection modeling \citep{Stellingwerf1982,SmolecMoskalik2008}.  In this work, \MESARSP{} is used as a controlled numerical laboratory: stellar parameters are tuned to match the period of Delta Cephei, and RSP damping parameters are then varied to diagnose nonlinear amplitude behavior.

This framing is important because Cepheid pulsation has a long observational and theoretical history. Cepheids entered the distance scale through the period--luminosity relation \citep{LeavittPickering1912,FreedmanMadore1991,Freedman2001,Pietrzynski2019}, while Delta Cephei itself has been the subject of earlier pulsation-modeling work \citep{Christy1966}. Modern nonlinear and convective pulsation calculations also build on decades of hydrodynamic model development \citep{Stellingwerf1982,SimonLee1981,SmolecMoskalik2008}. More recently, \MESA{} and \MESARSP{} have been used in broad Cepheid grids, multi-band light-curve comparisons, and RSP parameter studies \citep{Paxton2019,Kurbah2023,Hocde2024,Deka2025,Smolec2026}. The novelty claimed here is therefore not that Cepheids or Delta Cephei have never been modeled; it is the explicit, reproducible Delta-Cephei workflow and the clean-versus-diagnostic acceptance logic applied to this particular model sequence.

The scientific problem addressed here is deliberately narrower than a complete observed-light-curve fit.  We do not attempt in this paper to present final synthetic multi-band photometry, detailed MESA Isochrones and Stellar Tracks (MIST) bolometric-correction results, or Fourier morphology optimization.  Those topics require additional choices about observational data cleaning, bolometric corrections, passband transformations, phase alignment, and Fourier metrics.  Instead, the present paper establishes the computational foundation needed by those later analyses: a corrected active-inlist workflow, a period-calibrated Delta Cephei RSP model, an amplitude-control sequence, and a transparent classification of clean, diagnostic, and rejected runs.

The central numerical result is that the tuned 200-period model reproduces the adopted period extremely well:
\begin{equation}
    P_{\rm obs}=5.366531~{\rm d},
    \qquad
    P_{\rm model}=5.366622~{\rm d}.
\end{equation}
The absolute difference is
\begin{equation}
    \Delta P = P_{\rm model}-P_{\rm obs}=9.1\times10^{-5}~{\rm d}
    \simeq 7.86~{\rm s},
\end{equation}
and the fractional period error is
\begin{equation}
    \frac{\Delta P}{P_{\rm obs}}\times100 \simeq 0.00170\%.
\end{equation}
However, the subsequent nonlinear evolution shows that such period agreement is not sufficient for accepting a model as a stable nonlinear solution.  This paper therefore emphasizes a key methodological lesson: period calibration and nonlinear model acceptance must be treated as separate steps.

The paper is organized as follows.  Section~\ref{sec:target} defines the adopted target period and modeling goals.  Section~\ref{sec:method} describes the reproducible \MESARSP{} workflow and diagnostic quantities.  Section~\ref{sec:periodcalibration} presents the tuned 200-period period-matched model.  Section~\ref{sec:periodnotenough} explains why the period match did not end the modeling process.  Section~\ref{sec:amplitudecontrol} presents the first amplitude-control sequence using \RSPalfam{}.  Section~\ref{sec:classification} defines the model classification scheme.  Sections~\ref{sec:discussion} and \ref{sec:conclusions} summarize the implications and next steps.

\subsection*{Scope and contribution}
This manuscript is written as a target-specific workflow and diagnostic study. It does not claim the first nonlinear model of Delta Cephei, the first use of \MESARSP{} for Cepheids, or a final observed-light-curve solution. Previous nonlinear Cepheid modeling and recent \MESARSP{} studies already provide broad Cepheid grids and light-curve comparisons \citep[e.g.,][]{Paxton2019,Kurbah2023,Hocde2024,Deka2026}. The narrower contribution here is to document a reproducible Delta-Cephei calculation sequence and to show, with explicit run classification, why a highly accurate period match can still be rejected or treated as diagnostic if the nonlinear amplitude behavior, surface velocities, or energy diagnostics are not acceptable.

The present paper also defines the nonlinear model sequence used by related companion studies. The observed Johnson-\(V\) benchmark and reproducible Fourier-template construction are presented separately \citep{ElahiGull2026AAVSOFourier}. A semi-empirical reconstruction using photometry, radial velocities, and temperature constraints is developed in a separate companion analysis \citep{ElahiGull2026SemiEmpirical}. The sensitivity of a fixed \dcep{} \MESARSP{} setup to native opacity choices is examined separately \citep{ElahiSirolaGull2026OpacitySensitivity}. These companion papers are cited to define the division of labor among the analyses; the present manuscript remains restricted to the \MESARSP{} setup, period calibration, amplitude-control sequence, and clean-versus-diagnostic model classification.

\section{Adopted Target and Modeling Goals}
\label{sec:target}

\subsection{Adopted period}
\label{subsec:adoptedperiod}

The primary observational calibration target for this paper is the pulsation period of Delta Cephei,
\begin{equation}
    P_{\rm obs}=5.366531~{\rm d}.
    \label{eq:pobs}
\end{equation}
This value is used throughout the period calibration and cycle-by-cycle diagnostic analysis.  The goal of the first modeling stage is to find a \MESARSP{} setup whose late-cycle period remains close to this value, while also producing a numerically interpretable nonlinear pulsation solution.

The use of the period as the first calibration target is physically motivated by the period--mean-density relation.  In its common form,
\begin{equation}
    P \simeq Q\left(\frac{R^3}{GM}\right)^{1/2},
    \label{eq:perioddensity}
\end{equation}
where \(Q\) is the pulsation constant, \(R\) is a representative stellar radius, \(M\) is the stellar mass, and \(G\) is the gravitational constant.  The luminosity--radius--temperature relation,
\begin{equation}
    L = 4\pi R^2 \sigma T_{\rm eff}^4,
    \label{eq:lrt}
\end{equation}
then provides practical guidance for parameter tuning: at fixed effective temperature, increasing luminosity increases the radius and tends to lengthen the radial pulsation period; at fixed radius, increasing mass tends to shorten the period.  In the present work, these relations were used only as calibration guides.  Final model acceptance was based on direct \MESARSP{} output and post-processing diagnostics rather than on equation~\eqref{eq:perioddensity} alone.

\subsection{Modeling goals}
\label{subsec:goals}

The main goals of this paper are:
\begin{enumerate}[leftmargin=2em]
    \item to construct a reproducible Delta Cephei \MESARSP{} workflow;
    \item to tune the model parameters until the adopted period is reproduced;
    \item to test whether the period-matched model remains acceptable when extended in time;
    \item to use the \RSPalfam{} damping parameter as a first amplitude-control experiment;
    \item to classify models as clean, reference, amplitude-enhanced, diagnostic, or rejected.
\end{enumerate}

This paper therefore treats \MESARSP{} models as nonlinear radial pulsation calculations with prescribed stellar parameters.  It does not claim that the full stellar evolutionary history of Delta Cephei has been uniquely recovered.  Evolutionary tracks, instability-strip crossings, and secular period change are separate problems and should be addressed with full evolutionary calculations in a later paper.

\subsection{Acceptance criteria}
\label{subsec:acceptancecriteria}

A model is not accepted solely because its period agrees with equation~\eqref{eq:pobs}.  The acceptance criteria used in this work are:
\begin{enumerate}[leftmargin=2em]
    \item the model must reach the intended run length, usually 300 or 500 pulsation periods;
    \item the late-cycle period must remain close to the adopted Delta Cephei period;
    \item the cycle-by-cycle radius-amplitude proxy must be physically interpretable;
    \item the surface-velocity diagnostic must remain controlled;
    \item the run must not produce a serious energy-conservation warning that compromises interpretation;
    \item the result must be reproducible from the documented active inlist and analysis workflow.
\end{enumerate}

These criteria deliberately separate a period-matched model from a final nonlinear pulsation solution.  A run may be scientifically useful as a diagnostic experiment even if it is not accepted as a clean model.

\section{Numerical Method and Reproducible Workflow}
\label{sec:method}

\subsection{MESA-RSP setup}
\label{subsec:mesasetup}

The calculations were performed with \MESA{} version r24.08.1 using the \MESARSP{} radial stellar pulsation capability.  The starting point was the Cepheid RSP test-suite style setup, which was then modified into a Delta Cephei-specific project directory.  The main tuned quantities were the stellar mass \(M\), effective temperature \(\Teff\), luminosity \(L\), hydrogen mass fraction \(X\), metallicity \(Z\), selected RSP damping/convection parameters, and the requested number of pulsation periods.

The RSP calculations are one-dimensional nonlinear radial pulsation calculations with time-dependent convection and damping parameters.  In this paper, the most important RSP control parameter is \RSPalfam{}, which is treated as an eddy-viscous damping control.  Larger values of \RSPalfam{} suppress nonlinear amplitude growth more strongly, while smaller values allow larger amplitude growth.  This interpretation is used operationally in the amplitude-control sequence described in Section~\ref{sec:amplitudecontrol}.  Other RSP parameters, including \RSPalfat{} and \RSPgammar{}, are important for broader calibration, but they are not the main focus of this paper.

\subsection{Corrected active-inlist workflow}
\label{subsec:inlistworkflow}

A key reproducibility issue was identified early in the project.  The working directory appeared to contain a modified Delta Cephei setup, but the run script could still call the original Cepheid test-suite header.  This meant that visual editing of an inlist was not sufficient to guarantee that the intended parameters were actually used in the calculation.

To fix this, a dedicated Delta Cephei active inlist and header were created:
\begin{center}
\code{inlist_rsp_DeltaCephei_active},
\qquad
\code{inlist_rsp_DeltaCephei_header}.
\end{center}
The run workflow was then checked to ensure that the active Delta Cephei setup, rather than the unmodified test-suite header, was being read.  This correction converted the project from an edited test case into a reproducible computational workflow.  The corrected active/header structure is therefore treated as part of the scientific method, not just as a code-management detail.

\subsection{Cycle-by-cycle diagnostics}
\label{subsec:diagnostics}

For each model, the nonlinear time series was analyzed cycle by cycle.  The principal diagnostics were:
\begin{enumerate}[leftmargin=2em]
    \item the cycle period \(P_n\), measured from successive pulsation cycles;
    \item a radius-amplitude proxy \(\Delta R_n\);
    \item the maximum surface velocity ratio \(\max(v_{\rm surf}/c_s)\), where \(c_s\) is the local sound speed;
    \item warning flags from the \MESA{} run, especially energy-error warnings.
\end{enumerate}

In practice, the cycle period can be represented as
\begin{equation}
    P_n = t_{n+1}^{\rm ref} - t_n^{\rm ref},
    \label{eq:cycleperiod}
\end{equation}
where \(t_n^{\rm ref}\) and \(t_{n+1}^{\rm ref}\) are successive reference phases, such as successive radius maxima.  The radius-amplitude proxy is
\begin{equation}
    \Delta R_n = R_{\max,n} - R_{\min,n},
    \label{eq:deltar}
\end{equation}
where the maximum and minimum are taken over the same pulsation cycle.  The surface-velocity diagnostic is useful because large or rapidly growing values may indicate that the model is entering a regime where shocks, numerical stiffness, or energy-conservation problems become important.

The diagnostic pipeline was used to decide whether an extended model remained stable, whether its amplitude was growing or saturating, and whether it could be promoted from a diagnostic experiment to a clean model.

\workflowfigure

\section{Period Calibration}
\label{sec:periodcalibration}

\subsection{Tuned 200-period model}
\label{subsec:tuned200}

The first successful Delta Cephei calibration was the tuned 200-period model listed in Table~\ref{tab:periodmatched}.  The adopted parameters were
\begin{equation}
    M=5.0~\Msun,\qquad \Teff=6050~{\rm K},\qquad L=2422.5~\Lsun,
\end{equation}
with composition
\begin{equation}
    X=0.73,\qquad Z=0.007.
\end{equation}
The model was evolved for 200 pulsation periods.  Post-processing of the nonlinear time series gave
\begin{equation}
    P_{\rm model}=5.366622~{\rm d}.
\end{equation}
Compared with the adopted observed period, this gives
\begin{equation}
    \Delta P = 9.1\times10^{-5}~{\rm d},
\end{equation}
which is approximately 7.86 s.  The fractional period difference is
\begin{equation}
    \epsilon_P = \frac{|P_{\rm model}-P_{\rm obs}|}{P_{\rm obs}}
    =1.70\times10^{-5},
\end{equation}
or \(0.00170\%\).

\begin{table}[htbp]
\centering
\caption{Tuned 200-period period-matched \MESARSP{} model of Delta Cephei.}
\label{tab:periodmatched}
\begin{tabular}{lc}
\toprule
Quantity & Value \\
\midrule
Mass & \(5.0~\Msun\) \\
Effective temperature & \(6050~{\rm K}\) \\
Luminosity & \(2422.5~\Lsun\) \\
Hydrogen mass fraction & \(X=0.73\) \\
Metallicity & \(Z=0.007\) \\
Run length & 200 periods \\
Adopted observed period & \(5.366531~{\rm d}\) \\
Measured model period & \(5.366622~{\rm d}\) \\
Absolute period difference & \(9.1\times10^{-5}~{\rm d}\) \\
Absolute period difference & \(7.86~{\rm s}\) \\
Fractional period error & \(1.70\times10^{-5}\) \\
Percentage period error & \(0.00170\%\) \\
\bottomrule
\end{tabular}
\end{table}

This result is the strongest period-calibration result of the first modeling stage.  It shows that the chosen \MESARSP{} parameter set can reproduce the observed period of Delta Cephei with very high precision.  However, the period match alone does not demonstrate that the nonlinear amplitude is stable, saturated, or observationally realistic.  The next step was therefore to extend and test the nonlinear behavior of the model.

\periodcalibrationfigure

\subsection{Physical interpretation of the period match}
\label{subsec:periodinterpretation}

The success of the 200-period calibration is consistent with the period--mean-density scaling in equation~\eqref{eq:perioddensity}.  By adjusting \(L\), \(\Teff\), and \(M\), the global mean-density structure can be brought into agreement with the observed radial pulsation period.  This does not require that the nonlinear pulsation amplitude or light-curve morphology be correct.  The period is therefore a necessary first constraint, but it is not a complete model-validation metric.

This distinction is especially important for Cepheid modeling because different combinations of stellar parameters and RSP damping parameters may yield similar periods while producing very different amplitude histories.  A model that is excellent in period can still be unsuitable as a nonlinear solution if it exhibits nonstationary amplitude growth, excessive surface velocities, or numerical energy problems.

\section{Why Period Matching Was Not Sufficient}
\label{sec:periodnotenough}

After the 200-period period-matched model was obtained, the calculation was extended to test whether the nonlinear behavior remained controlled.  In the default or weak-damping extension, the model did not provide a satisfactory final solution.  The amplitude and cycle behavior became nonstationary, and the model could not be accepted solely on the basis of the earlier period agreement.

This result is the main methodological turning point of the paper.  The 200-period model answered the question, ``Can the period be matched?''  It did not answer the more difficult question, ``Does the nonlinear model remain physically and numerically acceptable when followed for many cycles?''  The answer to the second question required additional diagnostics.

The nonstationary extension showed that the problem was no longer primarily a mean-density tuning problem.  Instead, it became a nonlinear amplitude-control problem involving the time-dependent convection and damping parameters in the RSP treatment.  Therefore, the modeling workflow was divided into two distinct stages:
\begin{enumerate}[leftmargin=2em]
    \item period calibration using global stellar parameters;
    \item amplitude-control testing using RSP damping/convection parameters.
\end{enumerate}

\begin{figure}[htbp]
\centering
\includegraphics[width=0.82\textwidth]{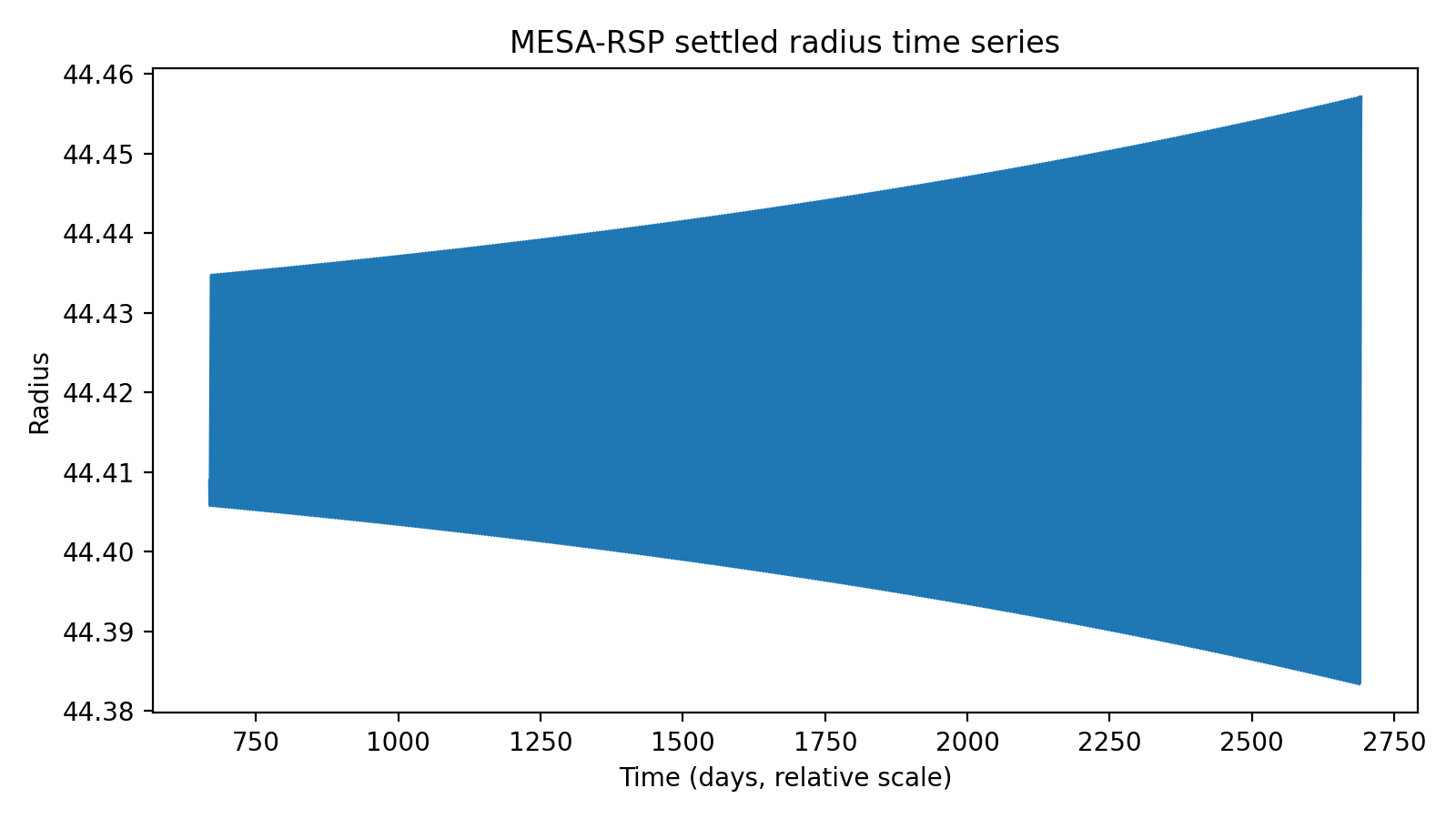}
\caption{Representative nonlinear radius time-series diagnostic from the Delta Cephei RSP runs. This figure supports the distinction between period agreement and nonlinear model behavior.}
\label{fig:radius_timeseries_reference}
\end{figure}

\section{Amplitude-Control Sequence with \RSPalfam{}}
\label{sec:amplitudecontrol}

\subsection{Role of \RSPalfam{}}
\label{subsec:alfamrole}

The first amplitude-control sequence focused on \RSPalfam{}, an RSP parameter that controls eddy-viscous damping in the nonlinear pulsation calculation.  Operationally, larger values of \RSPalfam{} suppress the pulsation amplitude more strongly, while smaller values allow stronger nonlinear amplitude growth.  This made \RSPalfam{} the natural first parameter to vary after the default extension showed nonstationary behavior.

The purpose of this sequence was not to claim a final observed-amplitude solution.  Instead, it was to determine whether the nonlinear amplitude could be increased while maintaining a period close to Delta Cephei and avoiding serious numerical warnings.

\subsection{Reference and amplitude-enhanced models}
\label{subsec:alfammodels}

Table~\ref{tab:alfamsequence} summarizes the main \RSPalfam{} sequence.  The \(\RSPalfam=0.60\), \(\RSPgammar=0.0\) model with \(L=2360~\Lsun\) is retained as the period-stable reference.  Its measured period is approximately
\begin{equation}
    P_{0.60}=5.368722~{\rm d}.
\end{equation}
This is slightly longer than the adopted observed period but sufficiently close to serve as a controlled reference for amplitude tests.

The intermediate \(\RSPalfam=0.50\) calculation increased the radius-amplitude proxy while keeping the period close to the target.  The \(\RSPalfam=0.425\), 500-period calculation is the most important result of this first amplitude-control stage.  It increased the nonlinear amplitude relative to the \(\RSPalfam=0.60\) reference while preserving a late-cycle period close to the observed value.  It is therefore classified as the clean amplitude-enhanced candidate from the first amplitude-control sequence.

\begin{table}[htbp]
\centering
\small
\caption{Main \RSPalfam{} amplitude-control sequence. All listed 500-period models use \(\RSPgammar=0.0\). The radius-amplitude proxy values are used only as internal nonlinear diagnostics; this paper does not claim a final observed-band amplitude solution.}
\label{tab:alfamsequence}
\begin{tabularx}{\textwidth}{>{\raggedright\arraybackslash}p{0.18\textwidth}c c >{\raggedright\arraybackslash}p{0.20\textwidth} X}
\toprule
\RSPalfam{} setting & Run length & Period (d) & Amplitude behavior & Classification \\
\midrule
0.60 & 500 periods & 5.368722 & smaller; controlled & Period-stable reference \\
0.50 & 500 periods & 5.367829 & intermediate & Amplitude-control comparison \\
0.425 & 500 periods & \(\simeq 5.367\) & larger & Clean amplitude-enhanced candidate \\
0.425 extension & 700 periods & close to target & continued growth & Diagnostic only; energy-error warning \\
\bottomrule
\end{tabularx}
\end{table}

\begin{figure}[htbp]
\centering
\includegraphics[width=0.92\textwidth]{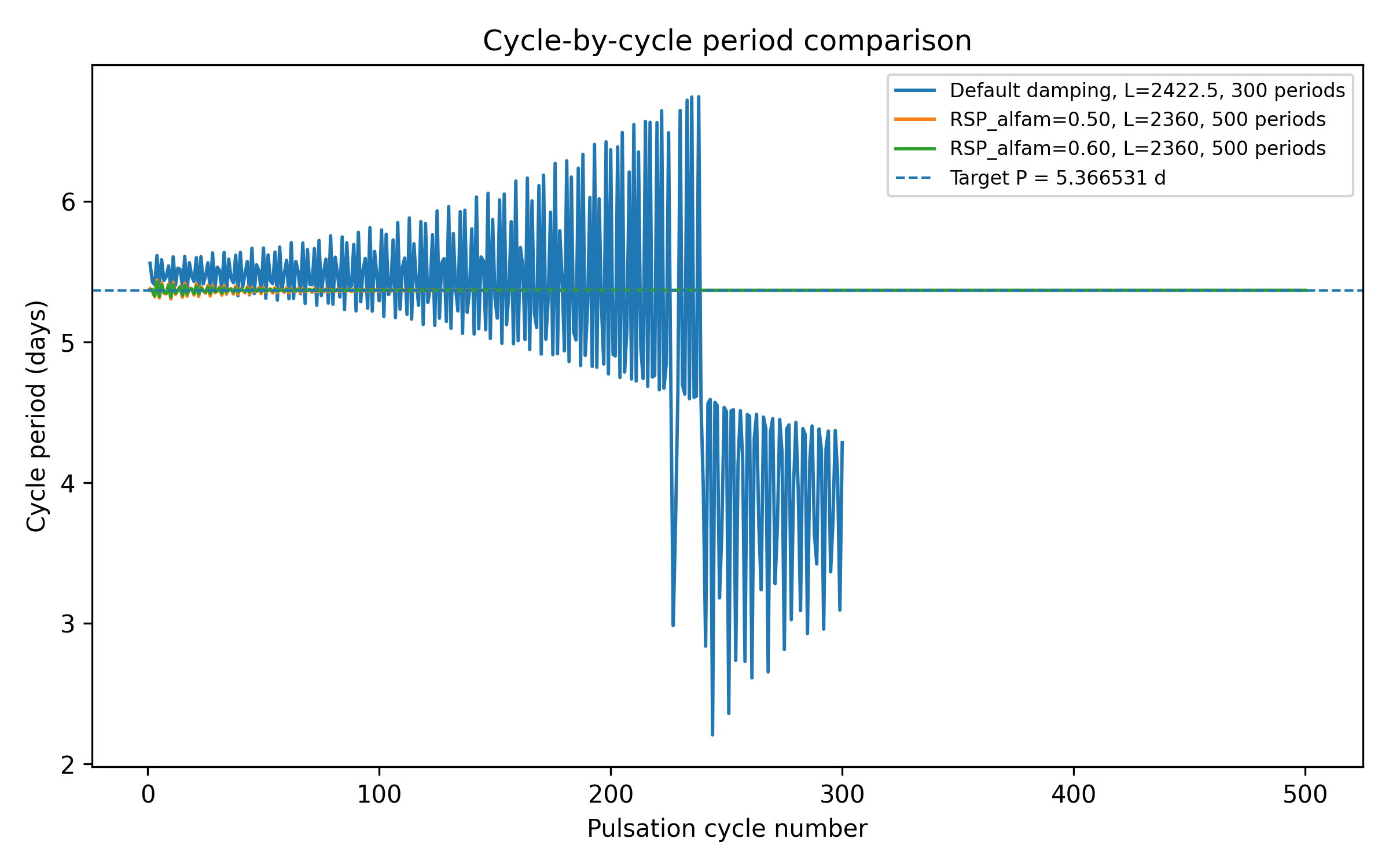}
\caption{Cycle-by-cycle period comparison for the main $\RSPalfam$ amplitude-control models. The figure is useful for showing that the amplitude-control sequence was evaluated not only by amplitude behavior, but also by whether the late-cycle period remained close to the Delta Cephei target.}
\label{fig:ampcontrol_period}
\end{figure}

\begin{figure}[htbp]
\centering
\includegraphics[width=0.92\textwidth]{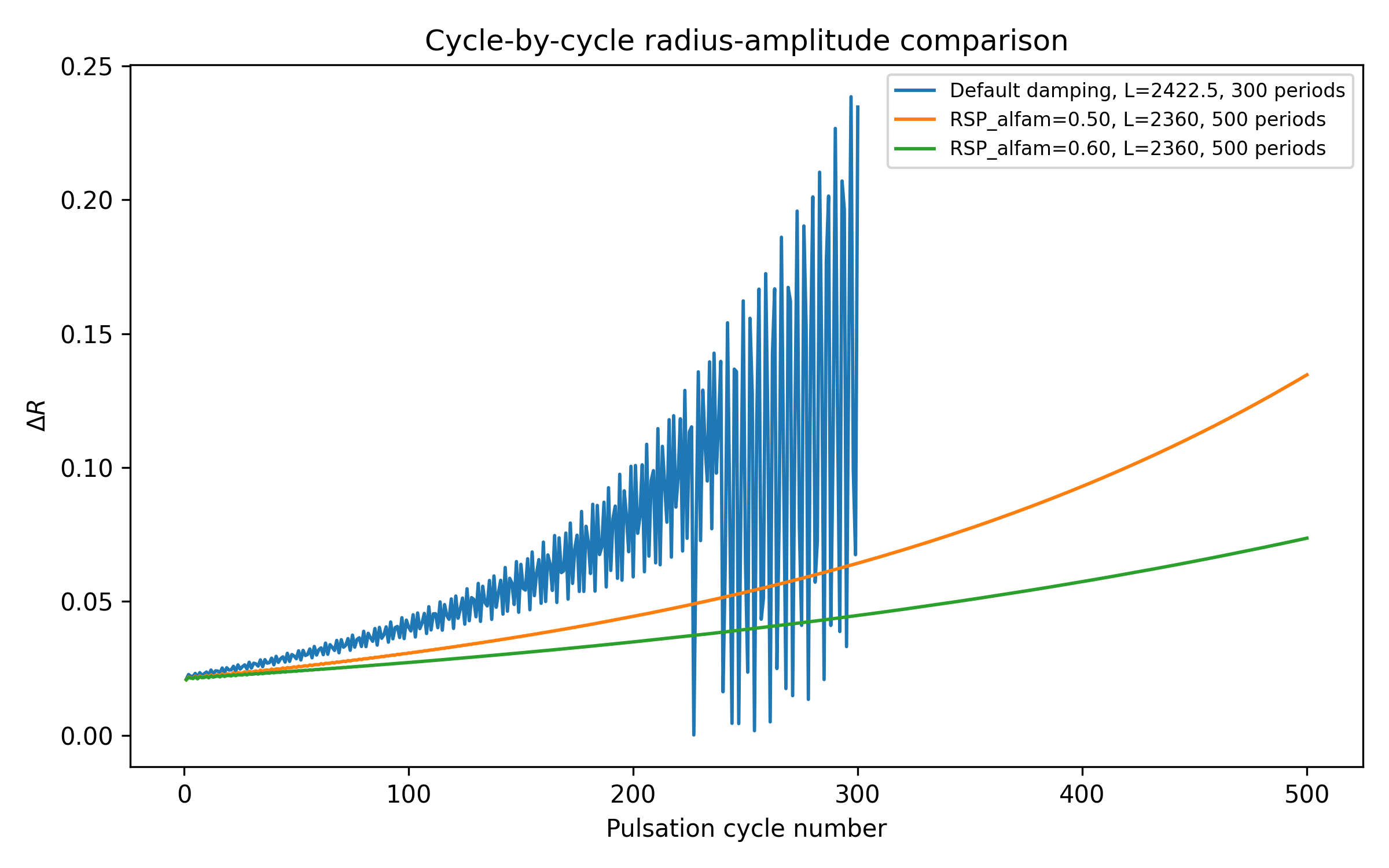}
\caption{Cycle-by-cycle radius-amplitude proxy for the $\RSPalfam$ amplitude-control sequence. Lower $\RSPalfam$ increases nonlinear amplitude growth, but clean model acceptance also requires controlled diagnostics and no serious energy-error warning.}
\label{fig:ampcontrol_deltar}
\end{figure}

\begin{figure}[htbp]
\centering
\includegraphics[width=0.92\textwidth]{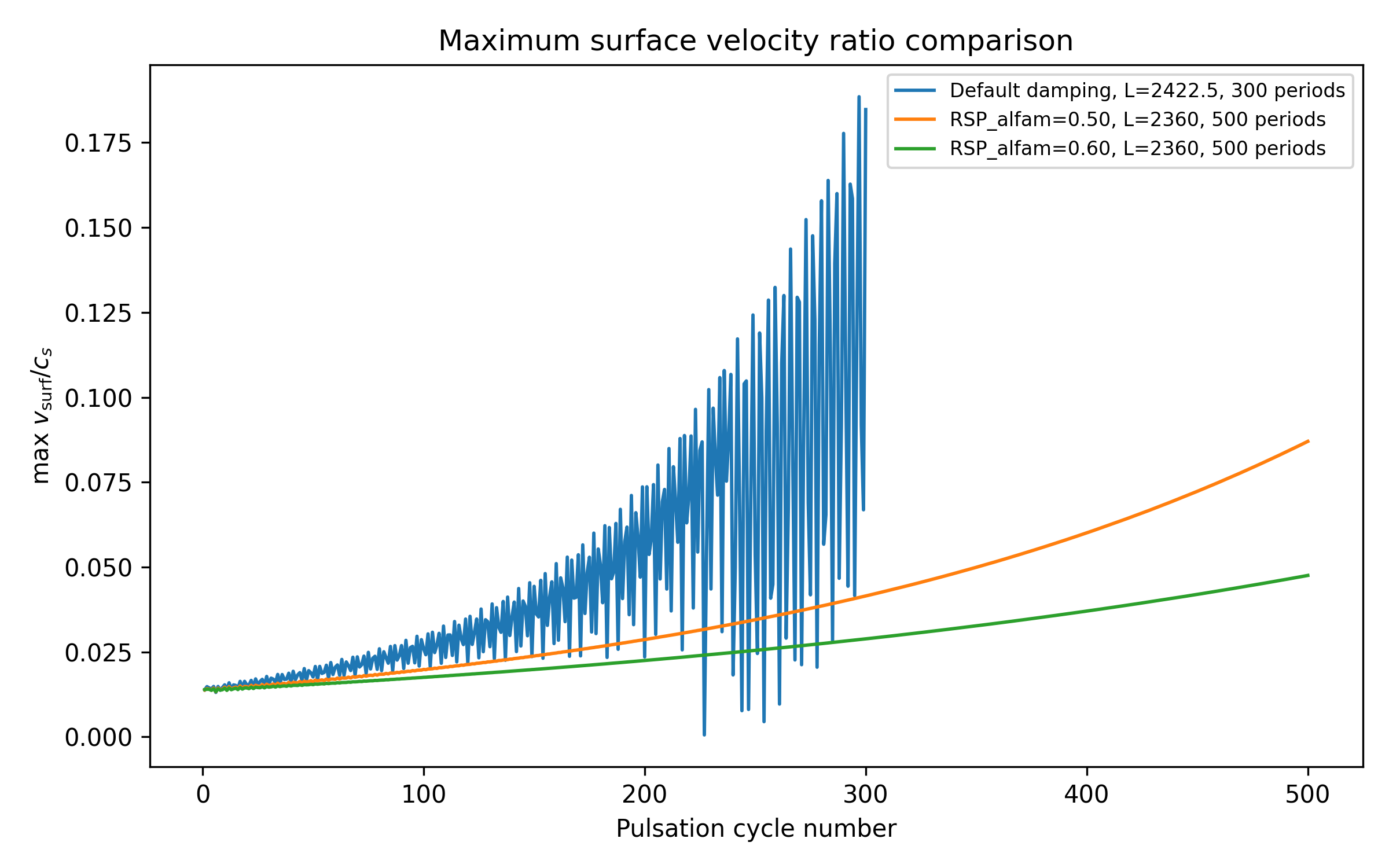}
\caption{Surface-velocity diagnostic for the $\RSPalfam$ sequence. This diagnostic helps distinguish useful amplitude growth from potentially problematic nonlinear behavior.}
\label{fig:ampcontrol_vsurf}
\end{figure}

\subsection{The 700-period extension}
\label{subsec:700extension}

The 700-period extension of the amplitude-enhanced case is scientifically informative but is not accepted as a clean final model.  It showed continued amplitude growth and retained a period close to the observed value, but it also produced a \MESA{} energy-error warning.  Because energy conservation is part of the acceptance criteria, this run is classified as a diagnostic growth experiment only.

This classification is important for avoiding overclaiming.  The 700-period extension demonstrates that lowering \RSPalfam{} can drive stronger nonlinear amplitude growth.  However, a model affected by an energy-error warning cannot be used as the final accepted physical solution without additional verification.  The clean model at this stage is therefore the \(\RSPalfam=0.425\), 500-period run, not the 700-period extension.

\section{Model Classification}
\label{sec:classification}

\subsection{Classification scheme}
\label{subsec:classificationscheme}

The model sequence naturally separates into clean models, reference models, amplitude-enhanced candidates, diagnostic models, and rejected models.  Table~\ref{tab:classificationdefs} defines these categories.  This classification scheme is useful because it preserves information from failed or warning-producing runs without allowing them to dominate the accepted scientific conclusions.

\begin{table}[htbp]
\centering
\caption{Model classification scheme used in this paper.}
\label{tab:classificationdefs}
\begin{tabular}{p{0.22\textwidth}p{0.40\textwidth}p{0.27\textwidth}}
\toprule
Class & Meaning & Use in this paper \\
\midrule
Clean model & Reaches requested run length, has interpretable late-cycle behavior, and has no serious numerical warning & Used for main conclusions \\
Reference model & Stable or controlled baseline used for comparison & Used to define period or amplitude baseline \\
Amplitude-enhanced candidate & Clean model with increased amplitude relative to the reference while preserving the period reasonably well & Used as best candidate from a calibration stage \\
Diagnostic model & Scientifically informative but affected by warnings, nonstationarity, or incomplete saturation & Discussed but not adopted as final \\
Rejected model & Incorrect setup, poor period behavior, serious numerical issue, or unsuitable nonlinear behavior & Listed for reproducibility only \\
\bottomrule
\end{tabular}
\end{table}

\subsection{Classification of the main models}
\label{subsec:classificationmain}

Table~\ref{tab:mainclassification} gives the classification of the principal models discussed in this paper.  The 200-period model is the period-matched proof of concept.  The \(\RSPalfam=0.60\) run is the period-stable reference for the amplitude-control stage.  The \(\RSPalfam=0.425\), 500-period run is the clean amplitude-enhanced candidate.  The 700-period extension remains diagnostic only.

\begin{table}[htbp]
\centering
\small
\caption{Classification of the main \MESARSP{} models discussed in this paper.}
\label{tab:mainclassification}
\begin{tabularx}{\textwidth}{>{\raggedright\arraybackslash}p{0.28\textwidth}>{\raggedright\arraybackslash}p{0.26\textwidth}>{\raggedright\arraybackslash}X}
\toprule
Model/case & Classification & Reason \\
\midrule
Corrected active/header setup & Reproducible workflow & Ensured that the Delta Cephei inlists, rather than the original test-suite header, were used \\
Tuned 200-period model & Period-matched proof of concept & Reproduced \(P_{\rm obs}\) with \(P_{\rm model}=5.366622~{\rm d}\) \\
Default/weak-damping extension & Diagnostic/rejected as final & Showed nonstationary nonlinear amplitude and cycle behavior \\
\(\RSPalfam=0.60\), 500 periods & Period-stable reference & Controlled amplitude and period near target \\
\(\RSPalfam=0.50\), 500 periods & Intermediate comparison & Increased amplitude relative to 0.60 while retaining period agreement \\
\(\RSPalfam=0.425\), 500 periods & Clean amplitude-enhanced candidate & Improved amplitude while preserving the period and avoiding serious warning \\
\(\RSPalfam=0.425\), 700-period extension & Diagnostic only & Continued amplitude growth but produced an energy-error warning \\
\bottomrule
\end{tabularx}
\end{table}

\classificationflowchart

\section{Discussion}
\label{sec:discussion}

\subsection{Period matching as a necessary but insufficient constraint}
\label{subsec:periodnecessary}

The tuned 200-period model demonstrates that \MESARSP{} can reproduce the adopted period of Delta Cephei with high precision.  This is an important success because the period is the first-order observable that any radial pulsation model must match.  The result also confirms that the chosen range of \(M\), \(L\), \(\Teff\), \(X\), and \(Z\) can generate a mean-density structure appropriate for Delta Cephei.

However, the later nonlinear behavior shows why period matching is not sufficient.  The default/weak-damping extension became nonstationary, and the amplitude behavior required separate calibration.  This is physically expected: the period depends strongly on the global mechanical structure, whereas nonlinear amplitude saturation depends on the balance of pulsational driving, damping, convection, and numerical stability.  Therefore, a period-matched model should be viewed as the beginning of nonlinear Cepheid modeling, not its endpoint.

\subsection{Amplitude control through RSP damping}
\label{subsec:amplitudedamping}

The \RSPalfam{} sequence provides a controlled first demonstration of how nonlinear amplitude behavior responds to RSP damping.  Increasing damping suppresses amplitude growth, while reducing damping allows larger amplitudes.  The \(\RSPalfam=0.60\) model is useful because it behaves as a stable reference.  The \(\RSPalfam=0.425\), 500-period model is useful because it increases the amplitude while maintaining an acceptable period and clean status.

The 700-period extension illustrates both the value and the risk of pushing toward larger amplitudes.  It shows continued growth and remains period-relevant, but the energy-error warning prevents it from being adopted as a clean model.  This emphasizes that high amplitude alone is not a sufficient criterion.  A model with larger amplitude but compromised energy behavior is less useful than a lower-amplitude model that remains clean and reproducible.

\subsection{Implications for future Delta Cephei modeling}
\label{subsec:implications}

The main implication is that Delta Cephei modeling should proceed in stages.  First, the period and reproducible workflow must be established.  Second, amplitude-control parameters should be explored systematically while applying strict numerical acceptance criteria.  Third, only clean candidates should be passed into synthetic photometry and Fourier morphology comparisons.  This staged approach prevents a warning-producing run from being mistaken for a final physical solution simply because it has a larger amplitude.

A broader calibration grid should eventually include additional RSP convection and damping parameters, including \RSPalfat{}, \RSPgammar{}, and related parameters that affect convective flux, turbulent transport, turbulent pressure, and radiative damping.  Such a grid should be evaluated not only by period and amplitude, but also by late-cycle stability, surface-velocity behavior, synthetic observed-band amplitudes, Fourier morphology, and residual structure against observed light curves.  Those later comparisons are beyond the scope of this paper.

\subsection{Limitations}
\label{subsec:limitations}

This paper has several intentional limitations.  First, the models are fixed-parameter RSP calculations rather than a full evolutionary reconstruction of Delta Cephei.  The stellar parameters are tuned to reproduce pulsation behavior, but the evolutionary crossing and secular period change are not solved here.  Second, the paper uses radius-amplitude and surface-velocity diagnostics rather than claiming a final observed-band amplitude solution.  Third, detailed synthetic photometry and Fourier morphology are deferred to companion analyses.  Fourth, the \RSPalfam{} sequence is only the first amplitude-control stage; it is not a full calibration of all RSP time-dependent convection parameters.

These limitations do not weaken the main result.  Rather, they define the proper scope of the paper: a reproducible, period-matched, nonlinear radial pulsation workflow with explicit amplitude-control diagnostics and model classification.

\section{Conclusions}
\label{sec:conclusions}

We have developed and tested a reproducible \MESARSP{} workflow for nonlinear radial pulsation modeling of Delta Cephei.  The main conclusions are:

\begin{enumerate}[leftmargin=2em]
    \item A corrected active-inlist/header workflow was required to ensure that the intended Delta Cephei setup was actually used.  The dedicated files \code{inlist_rsp_DeltaCephei_active} and \code{inlist_rsp_DeltaCephei_header} form the reproducible basis of the model sequence.

    \item A tuned 200-period \MESARSP{} model with \(M=5.0~\Msun\), \(\Teff=6050~{\rm K}\), \(L=2422.5~\Lsun\), \(X=0.73\), and \(Z=0.007\) reproduced the adopted Delta Cephei period extremely accurately, giving \(P_{\rm model}=5.366622~{\rm d}\) compared with \(P_{\rm obs}=5.366531~{\rm d}\).

    \item The period difference is \(9.1\times10^{-5}~{\rm d}\), or approximately 7.86 s.  The fractional period error is \(1.70\times10^{-5}\), corresponding to \(0.00170\%\).

    \item The period-matched model did not by itself solve the nonlinear modeling problem.  When extended with the default/weak damping configuration, the model showed nonstationary amplitude and cycle behavior.  Therefore, period calibration and nonlinear model acceptance must be treated as separate steps.

    \item The \(\RSPalfam=0.60\), \(\RSPgammar=0.0\) model is retained as the period-stable reference for the amplitude-control stage.

    \item The \(\RSPalfam=0.425\), 500-period model is identified as the clean amplitude-enhanced candidate from the first amplitude-control sequence.  It improves nonlinear amplitude behavior relative to the \(\RSPalfam=0.60\) reference while preserving a period close to the observed value.

    \item The 700-period extension is treated as diagnostic only because it produced a \MESA{} energy-error warning.  It is useful for understanding amplitude growth, but it is not adopted as a clean final model.

    \item The main methodological result is that a Cepheid model can match the observed period and still fail as a nonlinear pulsation solution.  Clean model acceptance requires period agreement, controlled late-cycle amplitude behavior, surface-velocity diagnostics, and no serious numerical energy warning.
\end{enumerate}

This paper establishes the modeling foundation for later work on observed-band synthetic photometry and Fourier morphology.  The next physical problem is a broader nonlinear amplitude calibration of the RSP convection and damping parameters, followed by consistent comparison with cleaned observed light curves.

\section*{Acknowledgments}

The authors acknowledge the developers of \MESA{} and the broader stellar pulsation community whose tools and methods made this work possible.

\section*{Data and Code Availability}

The calculations were performed with \MESA{} version r24.08.1 using a Delta Cephei-specific \MESARSP{} active inlist and header.  The primary model outputs are the \code{history.data} files and associated RSP diagnostic products.  The post-processing scripts extract cycle periods, radius-amplitude proxies, surface-velocity diagnostics, and warning/status information.  The inlists, analysis scripts, and reduced diagnostic tables are available from the corresponding author upon reasonable request.

\appendix
\section{Expanded Model Inventory for Reproducibility}
\label{app:modelinventory}

Table~\ref{tab:expandedinventory} gives a compact reproducibility inventory for the principal model sequence discussed in the main text.

\begin{table}[h]
\centering
\small
\caption{Compact reproducibility inventory of the principal models.}
\label{tab:expandedinventory}
\begin{tabularx}{\textwidth}{>{\raggedright\arraybackslash}p{0.21\textwidth}>{\raggedright\arraybackslash}X>{\raggedright\arraybackslash}p{0.22\textwidth}>{\raggedright\arraybackslash}p{0.20\textwidth}}
\toprule
Model/case & Main setup & Outcome & Decision \\
\midrule
Cepheid test-suite template & Original \MESARSP{} Cepheid test case & Initial working template & Starting point only \\
Wrong-header case & Edited directory could still call original header & Reproducibility problem identified & Rejected; motivated active/header correction \\
Corrected active/header case & dedicated active inlist and header & Correct Delta Cephei setup used & Reproducible workflow \\
Tuned 200-period model & \(M=5.0~\Msun\), \(\Teff=6050~{\rm K}\), \(L=2422.5~\Lsun\), \(X=0.73\), \(Z=0.007\) & \(P=5.366622~{\rm d}\) & Period-matched proof of concept \\
Default/weak-damping extension & Extension after period match & Nonstationary amplitude/cycle behavior & Diagnostic; not final \\
\(\RSPalfam=0.60\) & \(L=2360~\Lsun\), \(\RSPgammar=0.0\), 500 periods & \(P=5.368722~{\rm d}\), controlled amplitude & Period-stable reference \\
\(\RSPalfam=0.50\) & 500 periods & \(P\simeq5.367829~{\rm d}\), larger amplitude than 0.60 & Intermediate comparison \\
\(\RSPalfam=0.425\) & 500 periods & \(P\simeq5.367~{\rm d}\), enhanced amplitude & Clean amplitude-enhanced candidate \\
\(\RSPalfam=0.425\) extension & 700 periods & Continued growth; energy-error warning & Diagnostic only \\
\bottomrule
\end{tabularx}
\end{table}

\section{Useful Diagnostic Definitions}
\label{app:diagnostics}

Cycle extraction was based on successive reference phases in the nonlinear time series.  The diagnostic definitions used in this work are:
\begin{align}
    P_n &= t_{n+1}^{\rm ref}-t_n^{\rm ref}, \\
    \Delta R_n &= R_{\max,n}-R_{\min,n}, \\
    S_n &= \max_n\left|\frac{v_{\rm surf}}{c_s}\right|.
\end{align}
Here \(P_n\) is the period of cycle \(n\), \(\Delta R_n\) is the radius-amplitude proxy, and \(S_n\) is the maximum surface-velocity ratio during the cycle.  A clean model should show a period close to the adopted target, controlled \(\Delta R_n\), controlled \(S_n\), and no serious energy warning.

\clearpage
\bibliographystyle{unsrtnat}
\bibliography{references}

@book{Cox1980,
  author    = {Cox, J. P.},
  title     = {Theory of Stellar Pulsation},
  publisher = {Princeton University Press},
  year      = {1980}
}

@article{Bono1999,
  author  = {Bono, G. and Caputo, F. and Castellani, V. and Marconi, M.},
  title   = {Classical Cepheid Pulsation Models},
  journal = {The Astrophysical Journal},
  volume  = {512},
  pages   = {711},
  year    = {1999}
}

@article{Bono2024,
  author  = {Bono, G. and Braga, V. F. and Pietrinferni, A.},
  title   = {Cepheids as Distance Indicators and Stellar Tracers},
  journal = {The Astronomy and Astrophysics Review},
  volume  = {32},
  pages   = {4},
  year    = {2024}
}

@article{Paxton2019,
  author  = {Paxton, B. and Smolec, R. and Schwab, J. and others},
  title   = {Modules for Experiments in Stellar Astrophysics: Pulsating Variable Stars, Rotation, Convective Boundaries, and Energy Conservation},
  journal = {The Astrophysical Journal Supplement Series},
  volume  = {243},
  pages   = {10},
  year    = {2019}
}

@article{Jermyn2023,
  author  = {Jermyn, A. S. and Bauer, E. B. and Schwab, J. and others},
  title   = {Modules for Experiments in Stellar Astrophysics ({MESA}): Time-Dependent Convection, Energy Conservation, Automatic Differentiation, and Infrastructure},
  journal = {The Astrophysical Journal Supplement Series},
  volume  = {265},
  pages   = {15},
  year    = {2023}
}

@article{Stellingwerf1982,
  author  = {Stellingwerf, R. F.},
  title   = {Convection in Pulsating Stars. I. Nonlinear Hydrodynamics},
  journal = {The Astrophysical Journal},
  volume  = {262},
  pages   = {330},
  year    = {1982}
}

@article{SmolecMoskalik2008,
  author  = {Smolec, R. and Moskalik, P.},
  title   = {Convective Hydrocodes for Radial Stellar Pulsation. Physical and Numerical Formulation},
  journal = {Acta Astronomica},
  volume  = {58},
  pages   = {193},
  year    = {2008}
}

@article{LeavittPickering1912,
  author  = {Leavitt, H. S. and Pickering, E. C.},
  title   = {Periods of 25 Variable Stars in the Small Magellanic Cloud},
  journal = {Harvard College Observatory Circular},
  volume  = {173},
  pages   = {1},
  year    = {1912}
}

@article{FreedmanMadore1991,
  author  = {Madore, B. F. and Freedman, W. L.},
  title   = {The Cepheid Distance Scale},
  journal = {Publications of the Astronomical Society of the Pacific},
  volume  = {103},
  pages   = {933},
  year    = {1991}
}

@article{Freedman2001,
  author  = {Freedman, W. L. and Madore, B. F. and Gibson, B. K. and others},
  title   = {Final Results from the Hubble Space Telescope Key Project to Measure the Hubble Constant},
  journal = {The Astrophysical Journal},
  volume  = {553},
  pages   = {47},
  year    = {2001}
}

@article{Pietrzynski2019,
  author  = {Pietrzy{\'n}ski, G. and Graczyk, D. and Gallenne, A. and others},
  title   = {A Distance to the Large Magellanic Cloud That Is Precise to One Per Cent},
  journal = {Nature},
  volume  = {567},
  pages   = {200},
  year    = {2019}
}

@article{Christy1966,
  author  = {Christy, R. F.},
  title   = {Pulsation Models of {Delta Cephei} and {Eta Aquilae}},
  journal = {The Astrophysical Journal},
  volume  = {145},
  pages   = {340},
  year    = {1966}
}

@article{SimonLee1981,
  author  = {Simon, N. R. and Lee, A. S.},
  title   = {The Structural Properties of Cepheid Light Curves},
  journal = {The Astrophysical Journal},
  volume  = {248},
  pages   = {291},
  year    = {1981}
}

@misc{Kurbah2023,
  author        = {Kurbah, K. and Deb, S. and Kanbur, S. M. and Das, S. and Deka, M. and Bhardwaj, A. and Randall, H. R. and Kalici, S.},
  title         = {A Multiphase Study of Theoretical and Observed Light Curves of Classical Cepheids in the Magellanic Clouds},
  year          = {2023},
  note          = {arXiv:2303.08393}
}

@article{Hocde2024,
  author  = {Hocd{\'e}, V. and Smolec, R. and Moskalik, P. and Singh Rathour, R. and Zi{\'o}{\l}kowska, O.},
  title   = {Pulsation Modelling of the Cepheid {Y Ophiuchi} with {RSP/MESA}},
  journal = {Astronomy \& Astrophysics},
  volume  = {683},
  pages   = {A233},
  year    = {2024}
}

@misc{Deka2025,
  author = {Deka, M. and Marconi, M. and Molinaro, R. and others},
  title  = {The Light Curve Model Fitting of Large Magellanic Cloud Cepheids: {MESA-RSP} versus Stellingwerf-Code Predictions},
  year   = {2025},
  note   = {arXiv:2510.26385}
}

@misc{Smolec2026,
  author = {Smolec, R. and Zi{\'o}{\l}kowska, O. and Singh Rathour, R. and Hocd{\'e}, V. and Wielg{\'o}rski, P.},
  title  = {Toward a Comprehensive Grid of Cepheid Models with {MESA}. III. Evolutionary and Pulsation Relations for Models with Core and Envelope Overshooting},
  year   = {2026},
  note   = {arXiv:2603.26111}
}

@article{Deka2026,
  author  = {Deka, M. and others},
  title   = {The Light Curve Model Fitting of Large Magellanic Cloud Cepheids: {MESA-RSP} versus Stellingwerf-Code Predictions},
  journal = {Astronomy \& Astrophysics},
  volume  = {705},
  pages   = {A186},
  year    = {2026}
}

@misc{ElahiSirolaGull2026OpacitySensitivity,
  author        = {Elahi, Z. and Sirola, C. and Gull, W.},
  title         = {Native-Opacity Sensitivity of a Fixed {Delta Cephei} {MESA-RSP} Pulsation Model},
  year          = {2026},
  archivePrefix = {arXiv},
  eprint        = {2607.02439},
  primaryClass  = {astro-ph.SR},
  note          = {arXiv:2607.02439 [astro-ph.SR]}
}

@misc{ElahiGull2026SemiEmpirical,
  author        = {Elahi, Z. and Gull, W.},
  title         = {Semi-Empirical Pulsation Reconstruction of {Delta Cephei} with Photometry, Radial Velocities, and Temperature Constraints},
  year          = {2026},
  archivePrefix = {arXiv},
  eprint        = {2606.29561},
  primaryClass  = {astro-ph.SR},
  note          = {arXiv:2606.29561 [astro-ph.SR]}
}

@misc{ElahiGull2026AAVSOFourier,
  author        = {Elahi, Z. and Gull, W.},
  title         = {A Reproducible {AAVSO} {Johnson-V} Fourier Template for the Prototype Cepheid {Delta Cephei}},
  year          = {2026},
  archivePrefix = {arXiv},
  eprint        = {2606.29543},
  primaryClass  = {astro-ph.SR},
  note          = {arXiv:2606.29543 [astro-ph.SR]}
}

\end{document}